\documentstyle[aps,psfig]{revtex}

\begin{document}

\draft

\title{
\hfill {\rm MKPH-T-99-03}\\
\vspace{0.5cm}
Analysis of Resonance Multipoles from 
Polarization Observables in Eta Photoproduction}
\author{L. Tiator, D. Drechsel and G. Kn\"ochlein}
\address{Institut f\"ur Kernphysik,
Johannes Gutenberg-Universit\"at,
D-55099 Mainz, Germany}
\author{C. Bennhold}
\address{Center for Nuclear Studies, Department of Physics, The George
Washington University, Washington, D.C., 20052}

\date{\today}
\maketitle

\begin{abstract}
  A combined analysis of new eta photoproduction data for total and
  differential cross sections, target asymmetry and photon asymmetry
  is presented. Using a few reasonable assumptions we perform the
  first model-independent analysis of the $E_{0+}$, $E_{2-}$ and
  $M_{2-}$ eta photoproduction multipoles.  Making use of the
  well-known $A_{3/2}$ helicity amplitude of the $D_{13}(1520)$ state
  we extract its branching ratio to the $\eta N$ channel, $\Gamma_{\eta
  N}/\Gamma = (0.08 \pm 0.01)\%$. At higher energies, we show that
  the photon asymmetry is extremely sensitive to small multipoles that
  are excited by photons in the helicity $3/2$ state. The new GRAAL
  photon asymmetry data at higher energy show a clear signal of the
  $F_{15}(1680)$ excitation which permits extracting an $F_{15}(1680)
  \rightarrow \eta N$ branching ratio of $(0.15^{+0.35}_{-0.10})\% $.
    

\end{abstract}

\section{INTRODUCTION}
Over the past years, eta photoproduction has demonstrated its
potential as a new, powerful tool to selectively probe certain
resonances that are difficult to explore with pions. It is well known
that the low energy behavior of the eta production process is governed
by the $S_{11}(1535)$ resonance \cite{Benn91,Tiat94,Benm95}. The
recent, precise measurements of total and differential cross sections
for eta photoproduction at low energies \cite{Krus95,Wilh93} have made
it possible to determine the $S_{11}(1535)$ resonance parameters with
unprecedented precision.  A well-known example of the power of the
$(\gamma, \eta)$ reaction is the extraction of the $A^{p}_{1/2}$
helicity amplitude of the $S_{11}(1535)$ state. Due to the combined
cusp-resonance nature of this resonance, analyses based solely on pion
photoproduction consistently underestimate this quantity with values
of about $60\cdot 10^{-3} GeV^{-1/2}$ \cite{Arndt96}, while
extractions from eta photoproduction result in numbers nearly twice as
high \cite{Krus95}. Recent coupled-channels analyses
\cite{sauermann97,feuster98} that properly include cusp as well as
resonance phenomena have confirmed a range of values consistent with
eta photoproduction.

However, because of the overwhelming dominance of the $S_{11}$ the
influence of other resonances in the same energy regime, such as the
$D_{13}(1520)$, is difficult to discern.  It has been pointed
out \cite{Tiat94} that polarization observables provide a new doorway
to access these non-dominant resonances by use of interference of the
dominant $E_{0+}$ multipole with the smaller multipoles.  In
particular, the polarized photon asymmetry was shown to be sensitive
to the $D_{13}(1520)$.  It is well-known that, in principle, for a
completely model-independent multipole analysis seven single and
double polarization observables have to be measured along with the
differential cross section for all isospin channels.  However, in
practice, the recent extraction of the small $E^{3/2}_{1+}$ multipole
at the $\Delta$(1232) energy \cite{Beck97} demonstrates that the use of
a few reasonable assumptions permits an almost model-independent
analysis with a restricted set of observables for a limited energy
range.

Recently, polarization data for the target and photon asymmetries in
eta photoproduction were measured at ELSA \cite{Bock98} and
GRAAL \cite{Hour98}, respectively, for the first time. Combining these
data with the unpolarized cross sections from MAMI, we have performed
an almost model-independent multipole analysis of the $l=0$ and 2 eta
photoproduction multipoles at threshold.  This permits a precise
determination of the $D_{13}(1520)$ contribution and an extraction of
new $D_{13}(1520)$ resonance parameters.

\section{MULTIPOLE ANALYSIS}

The three measured observables are represented by the response
functions of \cite{Knoe95} as follows:
\begin{eqnarray}
\frac{d \sigma}{d \Omega}& = &\frac{q_{\eta}}{k}
R_T^{00}\, ,\\
T & = & \frac{R_T^{0y}}{R_T^{00}}\, ,\\
\Sigma & = & - \frac{^{c}R_{TT}^{00}}{R_T^{00}}\, ,
\end{eqnarray}
where $q_\eta$ and $k$ are the absolute values of the eta and photon
momenta, respectively, and here and in the following all variables are
expressed in the $cm$ frame.

Because of the overwhelming dominance of the $S_{11}$ channel in eta
photoproduction, these observables can be expressed in terms of
$s$-wave multipoles and interferences of the $s$ wave with other
multipoles. In the CGLN basis this leads to an $F_1$ dominance and the
observables can simply be expressed as
\begin{eqnarray}
\label{cgln}
R_T^{00} & = & |F_1|^2 - \mbox{Re} \left\{ 2 \cos\theta F_1^* F_2 -
    \sin^2\theta F_1^*F_4 \right\},\\
R_T^{0y} & = & 3 \sin\theta\, \mbox{Im} \left\{ F_1^* F_3 +
    \cos\theta F_1^*F_4 \right\},\\
^cR_{TT}^{00} & = & \mbox{Re} \left\{F_1^* F_4 \right\},
\end{eqnarray}
where $\theta$ is the scattering angle. If we retain only
interferences with $p$ and $d$ waves (an approximation valid at least
up to 1 GeV photon $lab$ energy) we obtain
\begin{eqnarray}
\label{o1}
R_T^{00} & = & |E_{0+}|^2 - \mbox{Re} \left[ E_{0+}^* 
\left( E_{2-} - 3 M_{2-} \right) \right] 
\nonumber \\
& & + 2 \cos \theta\, \mbox{Re} \left[ E_{0+}^* 
\left( 3 E_{1+} + M_{1+} - M_{1-} \right) \right]
\nonumber \\
& & + 3 \cos^2 \theta\, \mbox{Re} \left[ E_{0+}^* 
\left( E_{2-} - 3 M_{2-} \right) \right]\, ,\\
\label{o2}
R_T^{0y} & = &
3 \sin \theta\, \mbox{Im} \left[ E_{0+}^* \left( E_{1+} - M_{1+} \right)
  \right] \nonumber \\
& & - 3 \sin \theta \cos \theta\, 
\mbox{Im} \left[ E_{0+}^* \left( E_{2-} + M_{2-} \right) \right]\, ,\\
\label{o3}
^{c}R_{TT}^{00}& = & 
- 3 \sin^2 \theta\, \mbox{Re} \left[ E_{0+}^* 
\left( E_{2-} + M_{2-} \right) \right] \, .
\end{eqnarray}
Using the angle-independent quantities
\begin{eqnarray}
a & = & |E_{0+}|^2- \mbox{Re} \left[ E_{0+}^*
\left( E_{2-}-3 M_{2-} \right) \right]\, ,\\
b & = & 2 \mbox{Re} \left[ E_{0+}^* 
\left( 3 E_{1+} + M_{1+} - M_{1-} \right) \right]\, ,\\
c & = & 3 \mbox{Re} \left[ E_{0+}^* 
\left( E_{2-} - 3 M_{2-} \right) \right]\, ,\\
d & = & 
\frac{3}{a + c/3}\mbox{Im} \left[ E_{0+}^* 
\left( E_{1+} - M_{1+} \right) \right]\, ,\\
e & = & 
-\frac{3}{a + c/3}
\mbox{Im} \left[ E_{0+}^* \left( E_{2-} + M_{2-} \right) \right]\, ,\\
f & = &  \frac{3}{a + c/3}
\mbox{Re} \left[ E_{0+}^* \left( E_{2-} + M_{2-} \right) \right]\, ,
\end{eqnarray}
we can express the observables by a power series in $\cos\theta$ that
can be fitted to the experimental data at various energies
\begin{eqnarray}
\frac{d \sigma}{d \Omega}& = & \frac{q_{\eta}}{k}
\left(a + b \cos \theta + c \cos^2 \theta\right)\, ,\\
T & = & \sin \theta \left(d + e \cos \theta \right)\, ,\\
\Sigma & = & f \sin^2 \theta \, .
\end{eqnarray}

Quite remarkable, a combined analysis of these three observables
allows a determination of the $d$-wave contributions to eta
photoproduction once the quantities $a$, $c$, $e$ and $f$ have been
determined from experiment. As was already pointed out in
Ref. \cite{Krus95}, the differential cross section alone determines the
magnitude of the s-wave multipole
\begin{equation}
| E_{0+} | =  \sqrt{a+\frac{c}{3}} = \sqrt{\frac{1}{4\pi}\frac{k}{q_\eta}
 \sigma_{total}} \,\, .
\end{equation}
With the knowledge of $e$ and $f$ the helicity $3/2$ multipole
$B_{2-}$, defined below, and the phase relative to the $S_{11}$
channel can be determined:
\begin{eqnarray}
| B_{2-} | \equiv | E_{2-} + M_{2-} | & = & 
    \frac{1}{3}\sqrt{(e^2+f^2)(a+c/3)}\, ,\\
\tan (\phi_{E_{0+}}-\phi_{B_{2-}}) & = & \frac{e}{f}\, .
\end{eqnarray}

As is well known the pion photoproduction $E_{1+}$ and $M_{1+}$
multipoles in the $\Delta(1232)$ region must have the same phase due
to the Watson theorem. For resonances at higher energies this relation
does
not hold anymore since other channels are open and background
rescattering can affect the phases of the electric and magnetic
multipoles in a different way. Neglecting such effects for now,
$\phi_{E_{\ell \pm}} = \phi_{M_{\ell \pm}} = \phi_{\ell \pm}$, we arrive at
\begin{eqnarray}
E_{\ell \pm} & = & | E_{\ell \pm} | e^{i \phi_{\ell \pm}} \, ,\\
M_{\ell \pm} & = & | M_{\ell \pm} | e^{i \phi_{\ell \pm}}\, ,
\end{eqnarray}
with the following expressions for the magnitudes of the $l=2$ multipoles:
\begin{eqnarray}
|A_{2-}|& = &\frac{1}{2}|3M_{2-}-E_{2-}|=-\frac{c}{6f}
   \sqrt{\frac{e^2+f^2}{a+c/3}} \,,\\ 
|E_{2-}|& = & \frac{1}{4} \sqrt{\left(a + \frac{c}{3}\right)
\left( e^2 + f^2 \right)}\; |1 + \frac{c}{3 f (a+c/3)}|  \, ,\\
|M_{2-}|& = & \frac{1}{12} \sqrt{\left(a + \frac{c}{3}\right)
\left( e^2 + f^2 \right)}\; |1 - \frac{c}{f (a+c/3)}|  \, .\\
\end{eqnarray}
It is obvious from the data \cite{Krus95} that the total cross section
can be perfectly fitted to a Breit-Wigner form in the region of the
$S_{11}(1535)$ resonance, which results in an $s$-wave dominated
differential cross section. An investigation of the background due to
the Born terms \cite{Tiat94} yielded a very small eta-nucleon
coupling constant, and this result was confirmed by more recent
coupled-channels analyses \cite{feuster98}. As a consequence, the
$E_{0+}$ multipole can, to a high degree of accuracy, be solely
described by the $S_{11}(1535)$ contribution parameterized through a
Breit-Wigner form \cite{Krus95}. The additional arbitrary phase of
the complex $E_{0+}$ multipole is usually set equal to zero by
convention. If one uses a Breit-Wigner parameterization of the complex
$E_{0+}$ multipole, the phase $\phi_{0+}$ is given by
\begin{equation}
\tan\phi_{0+}(W) =  \frac{\Gamma(W) M_R}{{M_R}^2 - W^2 }\, ,
\end{equation}
where $W$ is the $cm$ energy and $M_R$ the mass of the resonance (conventional
resonance position). The energy dependent width of the
resonance is given by
\begin{equation}
\Gamma(W) = \Gamma_R \left( 
b_{\eta} \frac{q_\eta}{q_{\eta,R}}
+ b_{\pi} \frac{q_\pi}{q_{\pi,R}} + b_{\pi\pi}
\right) \, ,
\end{equation}
where $b_{\eta}, b_{\pi}$ and $b_{\pi\pi}$ denote the branching
ratios into the $\eta N$, $\pi N$ and $\pi \pi N$ channels,
respectively.

The analysis of the interference between the $E_{0+}$ and the $E_{2-}$
and $M_{2-}$ multipoles determines the $d$-wave multipoles and
therefore the difference $\phi_{2-} - \phi_{0+}$. It does not yield
direct information on $\phi_{2-}$. However, using the above
assumptions for the $E_{0+}$ multipole we can then obtain the phase
$\phi_{2-}$. Alternatively, if we assume a Breit-Wigner shape for the
$D_{13}$ resonance multipoles, we can obtain the phase $\phi_{0+}$.

To perform a similar analysis of the $p$-wave multipoles requires more
information from additional polarization observables; in particular, a
measurement of the recoil polarization would be useful. As before, we
obtain
\begin{equation}
P  = \frac{R_T^{y0}}{R_T^{00}} 
  = \sin \theta \left(h + k \cos \theta \right)\,
\end{equation}
with
\begin{eqnarray}
h & = & - \frac{1}{a + c/3} \mbox{Im} 
\left[ E_{0+}^* \left( 2 M_{1-} + 3 E_{1+} + M_{1+} \right) \right]\, ,\\
k & = & 3 \frac{1}{a + c/3}
\mbox{Im} \left[ E_{0+}^* \left( E_{2-} - 3 M_{2-} \right) \right]\, .
\end{eqnarray}

In analogy to Eqs. (19, 20) we can determine the
helicity 1/2 multipole $A_{2-}$ of the $D_{13}$ channel in a
model-independent way,
\begin{eqnarray}
| A_{2-} | \equiv \frac{1}{2} | 3 M_{2-} - E_{2-} | 
& = & \sqrt{\frac{c^2 + (a+c/3)^2 k^2}
{a+c/3}}\, ,\\
\tan (\phi_{A_{2-}}-\phi_{E_{0+}}) & = & \frac{(a+c/3)k}{c}\, .
\end{eqnarray}
Furthermore, together with Eqs. (11 and 13), the quantity $h$ allows
one to determine the resonance structure of the $p$-wave multipoles.

\section{RESULTS}

\subsection{Extraction of the multipoles}

Fig. 1 shows 4 out of 10 angular distributions measured by the TAPS
collaboration at Mainz \cite{Krus95} in the energy range between 716
and 790 MeV. While the isobar model of Ref. \cite{Knoe95} falls somewhat
low near threshold, a perfect fit is possible using the ansatz of
Eq. (16). The coefficient $a$ can be fitted to
a Breit-Wigner form with an energy-dependent width leading, e.g., to
$M_R=(1549\pm 8)MeV$, $\Gamma_R=(202\pm 35) MeV$ and an
absolute value of the $s$-wave multipole at threshold, $|E_{0+}|=16
\cdot 10^{-3}/m_\pi^+$ (Fit 1, Ref.~\cite{Krus95}). For our present
purpose, however, 
it is more convenient to use a general polynomial expansion for
the coefficients as described in Section III.C.

Fig. 2 shows our fit to the target polarization data from Bonn
\cite{Bock98}. In this case the isobar model of Ref. \cite{Knoe95}
fails to reproduce the angular shape of the data. In particular, the
model does not reproduce the node found experimentally at low energy.
Furthermore, the model ingredients are quite insensitive to the
$D_{13}$ resonance.  In previous coupled-channels analyses
\cite{Benn91,Tiat94} the $D_{13}$ resonance came out much stronger and
a node developed. However, the node resulted in a negative asymmetry
at forward and a positive asymmetry at backward angles, clearly
opposite to the experimental observation and, as we shall see below,
leading to a drastically different relative phase between $s$ and $d$
waves. Quite to the contrary, the ansatz of Eq.~(17) does describe the
data and leads to a node at energies below 800 MeV.

Fig. 3 compares our fit and the isobar model of Ref. \cite{Knoe95} for
the photon asymmetry. This observable has recently been measured at
GRAAL \cite{Hour98} from threshold up to 1050 MeV. At the lower
energies, the good agreement between the data and the isobar model
illustrates the importance of the $D_{13}$(1520) resonance.  Without
the $D_{13}$ the polarized photon asymmetry is almost zero for
energies below 900 MeV and turns negative for the higher energies.
With regard to our multipole analysis, Fig. 3 clearly demonstrates
that we can achieve an excellent fit with the ansatz of Eq.~(18). Up
to 900 MeV the asymmetry has a clean $\sin^2 \theta$ dependence and
can be parameterized by a single energy-dependent parameter $f$.
However, above 900 MeV the data show the evolution of a
forward-backward asymmetry that becomes most pronounced at 1050 MeV.
This behavior cannot be fitted any longer with the form of Eq.~(18)
but requires an additional coefficient,
\begin{eqnarray}
\Sigma & = & \sin^2 \theta (f + g \cos \theta) \, ,
\end{eqnarray}
where $g$ is determined solely by multipoles of order 3 and higher,
\begin{eqnarray}
  g & = & \frac{15}{a + c/3}
\mbox{Re} \left[ E_{0+}^* \left( E_{3-} + M_{3-} +  
E_{3+} - M_{3+}\right) \right]\, .
\end{eqnarray}

The obvious need for the coefficient $g$ at higher energies therefore
represents a clear signal that partial waves beyond $d$ waves are
required by the photon asymmetry data.

\subsection{The photon asymmetry at higher energies}

As has been discussed above, eta photoproduction at low energies is
dominated by $s$ waves giving rise to essentially flat angular
distributions with only small modulations as found by the Mainz
precision experiment \cite{Krus95}. But also the angular distributions
measured at Bonn up to $1.15$~GeV \cite{Bock97} have given no evidence
for a break-down of this $s$-wave dominance. This gives us the
possibility to extract the small contributions of the higher
resonances in exactly the same way as shown above for the $D_{13}$
resonance, i.e. by analyzing the interference with the leading
$s$-wave multipole.

In the following we shall demonstrate this method for the nucleon
resonances with strong helicity $3/2$ couplings $A_{3/2}$.

Assuming $s$-wave dominance, and therefore, $F_1$-dominance in the
CGLN basis, we can derive a general expression for the photon
asymmetry by using Eq. (6),
\begin{eqnarray}
\Sigma(\theta) & = & -\sin^2\theta\,\, 
\mbox{Re}\big[F_1^* F_4\big]/R_T^{00}\, \\
& = & \sin^2\theta\,\, \mbox{Re}\bigg[E_{0+}^* \sum_{\ell \ge 2}
(B_{\ell -}+B_{\ell +}) P_\ell''(\cos\theta) \bigg] /R_T^{00}\,\nonumber
\end{eqnarray}
with $B_{\ell-}=E_{\ell-}+M_{\ell-}$ and
$B_{\ell+}=E_{\ell+}-M_{\ell+}$.

Both multipole combinations have helicity $3/2$ and, for resonance
excitation, are proportional to the photon couplings $A_{3/2}$. The
helicity $1/2$ couplings $A_{1/2}$ do not enter in Eq.~(37) but only
appear in the differential cross section and in the recoil
polarization, e.g., as $A_{2-}=(3M_{2-}-E_{2-})/2$.  Expanding to
$\ell_{max}=4$, we then obtain
\begin{eqnarray}  
\Sigma(\theta) & = & \frac{\sin^2\theta}{|E_{0+}|^2}\, \mbox{Re}
\Big\{E_{0+}^*\Big[3(B_{2-}+B_{2+})-\frac{15}{2}(B_{4-}+B_{4+})
\nonumber \\
& + & 15(B_{3-}+B_{3+})\cos\theta 
+ \frac{105}{2}(B_{4-}+B_{4+})\cos^2\theta \Big] \Big\}\, .
\end{eqnarray}

This result demonstrates that any deviation from the $\sin^2 \theta$
dependence in the photon asymmetry is due to $f$ waves or higher
partial waves.  These could originate from either strong background
contributions or resonances with spin 5/2 or higher. We point out that
such contributions would appear in the differential cross section as
$\cos^3 \theta$ terms, which would be difficult to extract. Since the
background contributions in eta photoproduction are known to be small,
the only remaining option for these partial waves is a resonance with
$J \ge 5/2$.

In Table \ref{tab1} we list all entries for $N^*$ resonances with isospin
$1/2$.  From this table one finds the $D_{13}$ as the strongest
candidate that contributes significantly to the photon asymmetry.  The
next higher-lying resonance with a strong helicity 3/2 coupling is the
$F_{15}(1680)$, which is known to play an important role in pion
photoproduction.  Since its mass matches the energy region of the
forward-backward asymmetry in the photon polarization, we conclude that
the GRAAL data reveal the presence of the $F_{15}(1680)$ resonance in
eta photoproduction. In section III.D we shall extract the $\eta N$
branching ratio from this signal.

At even higher energies and beyond our present scope of interest,
there are the less-established $F_{17}(1990)$ and $G_{17}(2190)$ whose
properties could be extracted by measuring the photon asymmetry at the
corresponding energies. We have verified that even if two small
resonances of different multipolarity would be excited in the same
energy region they will produce a clear signal allowing us to
determine the $\eta$ branching ratios down to values well below
$0.1\%$. In Fig. 4 we demonstrate how such interferences of higher
resonances with the $S_{11}$-channel would show up in the photon
asymmetry.

\subsection{Parameterization of the multipoles}

Fig. 5 compares the results of our multipole analysis with the isobar
model calculation of Ref. \cite{Knoe95}.  The most dramatic difference
occurs for the relative phase between the $s$ and $d$ waves. As shown
in Eq.~(21), this phase difference is model independent. If both the
$S_{11}(1535)$ and the $D_{13}(1520)$ are parameterized by Breit-Wigner
functions (as in the case of the isobar model \cite{Knoe95}), this
phase difference would be rather constant, because both resonances are
very close in energy and, furthermore, have similar resonance widths.
Since the $S_{11}$ is a little bit higher in energy, the phase
difference $\phi_0 - \phi_2$ should be small and negative as shown by
the dotted line in Fig. 5.

From the above analysis we conclude that this unexpected discrepancy
is directly connected to the node structure in the target asymmetry.
Without a node or with a node but an $e$ coefficient of opposite sign,
the phase difference would be much smaller and closer to our model
calculations. Other models based on either a coupled-channels approach
\cite{feuster98} or on a tree-level effective Lagrangian analysis
\cite{mathur98}, have similar problems to reproduce the target
asymmetry. It is therefore imperative that this measurement be
verified as soon as possible.
 
After performing single-energy fits we used a polynomial form for the
energy dependence of the coefficients $a$, $b$, $c$, $d$, $e$ and $f$
of Eqs. (10)-(15) in order to arrive at a global (energy-dependent)
solution for the multipoles. This has several advantages: First, the
experimental data were obtained in different set-ups at different
labs, thus their energy bins do not match. Second, except for the
quantity $a$ that can be determined already from the total cross
section, all other quantities contain considerable error bars.
Therefore, a combined fit can reduce the uncertainty of individual
measurements considerably. In a simple Taylor expansion in terms of
the eta momentum with only 1 to 3 parameters in each coefficient, we
obtain good results for the energy region from threshold up to about
900 MeV. Using polynomials as a function of $q_{\eta} / m_\eta$ as
shown in Table \ref{tab2}, we obtain very good fits in the complete
energy range of the present experiments.  In Table \ref{tab2} we made
use of the constraints for the threshold behavior of the multipoles,
in particular the vanishing of the $p$-, $d$- and $f$-wave multipoles
at threshold. In Figs. 6 and 7 we show our energy dependent fits to
the coefficients $a$ to $g$ obtained at single energies. The
coefficients of the differential cross section in Fig. 6 reproduce the
findings of Krusche et al. \cite{Krus95}, the dominance of the
$s$-wave amplitude in the coefficient $a$, an $s$-$p$ interference
compatible with zero in the coefficient $b$, and a small $s$-$d$
interference in the coefficient $c$. Fig. 7 shows the coefficients $d$
to $g$ obtained from fits to the target and photon asymmetries. In
this case the coefficient $e$ is not very well defined by the target
asymmetry and our energy dependent fit gives just the most likely
polynomial description. Finally, the strong rise in the coefficient
$g$ around $1\,GeV$ shows the excitation of the $F_{15}(1680)$
resonance, which was measured just up to the resonance peak expected
at $E_{\gamma}^{lab}=1035\,MeV$. An experiment at even higher energies
is now in progress at GRAAL with the aim to confirm the resonance
behavior.

\subsection{Helicity amplitudes and branching ratios}

With the definitions given in Appendix A we are now in a position to
determine the helicity amplitudes and/or branching ratios.  We present
two separate analyses of resonance parameters based on two different
assumptions: First, we perform a model-independent analysis that uses
both the target asymmetry with coefficient $e\cos\theta\sin\theta$ and
the beam asymmetry with $f\sin^2\theta$. Second, because of the fact
that the phase difference $\Delta\phi=\phi_0 - \phi_2$ is mainly due
to the target asymmetry, we also perform an analysis ignoring the
measurement of $T$ and assuming a normal resonance behavior for the
$S_{11}(1535)$, with a phase given by the Breit-Wigner form, Eq. (28).

According to appendix A the branching ratio is related to the
photoproduction amplitude $B_{2-}$ by
\begin{equation}
\sqrt{b_\eta}=\left( \frac{3 \pi q_{\eta,R} M_R \Gamma_R}{k_R
    m_N}\right)^{1/2} \frac{\widetilde{B}_{2-}}{A_{3/2}}\, ,
\end{equation}
where we have introduced $b_\eta = \Gamma_\eta / \Gamma_R$.
In case 1 the amplitude $\widetilde{B}_{2-}$ is evaluated with
Eq. (20), in case 2 we use
\begin{equation}
\left.{\widetilde{B}_{2-}}=\frac{\sqrt{a+c/3}}{\sin\delta_0}\; 
\frac{f}{3} \right|_{W=1520 MeV} \,    
\end{equation}
with $\delta_0 = 72^\circ$ in contrast to $136^\circ$ from our
model-independent analysis (Fig. 4).
Using Eq. (24) to evaluate the helicity 1/2 amplitude
$\widetilde{A}_{2-}$ we can now determine the helicity ratio 
\begin{equation}
\frac{A_{3/2}}{A_{1/2}}=-\frac{\sqrt{3}}{2}\frac{\widetilde{B}_{2-}}
  {\widetilde{A}_{2-}} =\sqrt{3}\frac{f}{c}(a+\frac{c}{3})\,,
\end{equation}
independent of the target asymmetry measurement. Furthermore, the
reduced photoproduction resonance parameters $\xi_i$, Eq. (A4),
are simply related to the amplitudes $A_{2-},B_{2-}$ by
\begin{equation}
\xi_{1/2}=-\sqrt{4\pi}\widetilde{A}_{2-}\quad \mbox{and}\quad
\xi_{3/2}=-\sqrt{3\pi}\widetilde{B}_{2-}\, .
\end{equation}
For our analysis of the $F_{15}(1680)$ resonance, there exists
experimental information only on the photon asymmetry and the total cross
section at $ 1\, GeV$. Therefore, only case 2 is applicable and we
obtain
\begin{eqnarray}
\sqrt{b_\eta} &=& \left( \frac{12 \pi q_{\eta,R} M_R \Gamma_R}{k_R
    m_N}\right)^{1/2} \frac{\widetilde{B}_{3-}}{A_{3/2}}\, ,\\
\widetilde{B}_{3-} &=& \left. \sqrt{\frac{k_R\sigma_{total}}{4\pi q_{\eta,R}}}
 \; \frac{g}{15\sin\delta_0} \right|_{W=1680 MeV} ,    
\end{eqnarray} 
with $\delta_0 = 137^\circ$ from Eq. (28). In the case of the $F_{15}$ resonance
we have no information on the helicity 1/2 amplitude $A_{3-}$ and
therefore $A_{1/2}$ is not determined. This helicity amplitude is only crudely known
from pion photoproduction.  The value for $\xi_{3/2}$ is obtained
from
\begin{equation}
\xi_{3/2}=\sqrt{12\pi}\widetilde{B}_{3-}\, .
\end{equation}

In Table III we show our numerical results for the resonance
parameters together with the values given by the Particle Data Tables
\cite{PDG98} and a recent analysis of Mukhopadhyay and Mathur
\cite{mathur98}. It is very interesting to note that though the
$D_{13}$ has a branching ratio of only $0.08\%$ according to our
analysis it nevertheless totally dominates the photon asymmetry as
shown in Fig. 3. The photon asymmetry would basically vanish up to
$E_\gamma^{lab}\approx 900\, MeV$ without this resonance contribution.
The branching ratio of the $F_{15}$ resonance is also found to be well
below $1\%$. Both resonances would have never been seen in the total
cross section, because the $S_{11}$ dominates by two orders of
magnitude, and even in the angular distribution very high-precision
data are required to observe the interferences with the $s$ wave.

Our present analysis is in remarkable agreement with the calculations
of Koniuk and Isgur \cite{Kon80} who predicted the $\eta N$ branching
ratios for the $S_{11}(1535)$, $D_{13}(1520)$ and $F_{15}(1680)$ to be
47\%, 0.09\% and 0.8\%, respectively, in the framework of the
constituent quark model with hyperfine quark-gluon interaction.  In
comparison with Mukhopadhyay and Mathur \cite{mathur98}, we confirm
their finding of a much smaller ratio $A_{3/2}/A_{1/2}$ for the
$D_{13}$ resonance than given in the Particle Data Tables and
predicted by the quark model.

\section{SUMMARY}

We have demonstrated that polarization observables are a powerful tool
in analyzing individual resonances in the eta photoproduction channel.
The strong dominance of the $S_{11}$ channel allows for a more
straightforward analysis than in the case of pion photoproduction.
Furthermore, the nonresonant background in eta physics is small due to
the weak coupling of the eta to the nucleon.  A combined analysis of
differential cross section, photon asymmetry and target polarization
yields an almost model-independent determination of $s$- and $d$-wave
multipoles. The target polarization measured at Bonn results in an
unexpected phase shift between the $S_{11}$ and $D_{13}$ resonances
that needs to be confirmed by further experiments. If this phase is
taken at face value, it implies that at least one of these resonances
(perhaps the $S_{11}$) behaves quite differently from regular nucleon
resonances and might even be a completely different phenomenon, as
suggested by Ref. \cite{hoehler97}.

In conclusion, future experiments on target polarization and photon
asymmetry are expected to solve another piece of the ``eta'' puzzle that
makes this field of physics so exciting and yield qualitative new
information on higher and less known structures. The photon asymmetry
recently measured at GRAAL clearly indicates a contribution of the
$F_{15}(1680)$ resonance with an $\eta N$ branching ratio of 0.15\%.
Similar experiments at even higher energies could help to pin down the
properties of less-established resonances like the $F_{15}(1990)$ and
the $G_{17}(2190)$.

\acknowledgments

We would like to thank G. Anton, J.-P. Didelez and D. Rebreyend for
very fruitful discussions. This work was supported by the Deutsche
Forschungsgemeinschaft (SFB201) and the US-DOE grant
DE-FG02-95-ER40907.

\appendix
\section{Helicity Matrix Elements}

Following the notation of Ref. \cite{Arndt90} the 
helicity matrix elements for nucleon resonance production can be written as

\begin{eqnarray}
A^{\ell+}_{\frac{1}{2}}&=& -\frac{C_m}{\alpha} \widetilde{A}_{\ell+}\,,\nonumber\\
A^{\ell-}_{\frac{1}{2}}&=&  \frac{C_m}{\alpha} \widetilde{A}_{\ell-}\,,\nonumber\\
A^{\ell+}_{\frac{3}{2}}&=&  \frac{C_m}{2\alpha}\sqrt{\ell(\ell+2)} 
     \widetilde{B}_{\ell+}\nonumber\,,\\
A^{\ell-}_{\frac{3}{2}}&=& -\frac{C_m}{2\alpha}\sqrt{(\ell-1)(\ell+1)} 
     \widetilde{B}_{\ell-}\nonumber\\
\end{eqnarray}
with 
\begin{eqnarray}
\alpha = \left( \frac{1}{\pi}\frac{k_R}{q_{m,R}}\frac{1}{(2J+1)}\frac{m_N}{M_R}
\frac{\Gamma_m}{\Gamma_R^2}\right)^{\frac{1}{2}}
\end{eqnarray}
and
\begin{displaymath}
        C_m = \left\{
                \begin{array}{lcl}
                -1&\,\mbox{for}\,&\eta N (I=\frac{1}{2})\\
                -\sqrt{3}&\,\mbox{for}&\pi N (I=\frac{1}{2})\\
                +\sqrt{\frac{2}{3}}&\,\mbox{for}&\pi N (I=\frac{3}{2})\\
                \end{array}     
              \right. 
\end{displaymath}

These relations allow the extraction of the helicity elements for a
resonance with mass $M_R$, total width $\Gamma_R$ and total spin $J$.
The partial decay width in the meson-nucleon decay channel $m$ is
given by $\Gamma_m$, and $k_R$, $q_{m,R}$ are the $cm$ momenta of the
photon and the meson at the resonance position. The helicity
multipoles $\widetilde{A}_{\ell\pm}$ and $\widetilde{B}_{\ell\pm}$ are
defined as

\begin{center}
\begin{tabular}{lll}
$\widetilde{A}_{\ell\pm}=\mbox{Im} A_{\ell\pm}|_{W=M_R}$ \quad & and &
  \quad $\widetilde{B}_{\ell\pm}=\mbox{Im} B_{\ell\pm}|_{W=M_R}$ \,,
\end{tabular}
\end{center}
and in general we have 
\begin{eqnarray}
A_{\ell+}&=&\frac{1}{2} [(\ell+2)E_{\ell+} + \ell M_{\ell+}]\,,\nonumber\\
A_{\ell-}&=&\frac{1}{2} [(\ell+1)M_{\ell-} + (\ell-1)E_{\ell-}]\,,\nonumber\\
B_{\ell+}&=&E_{\ell+} - M_{\ell+}\nonumber\,,\\
B_{\ell-}&=&E_{\ell-} + M_{\ell-}\nonumber \,.
\end{eqnarray}

As introduced by Mukhopadhyay et al. \cite{mukho95} we can express the
photoproduction multipole at resonance by the quantities
\begin{eqnarray}
\xi_i&\equiv& \sqrt{\frac{m_N k_R \Gamma_m}{M_R q_{m,R} \Gamma_R^2}}
\, A_i\,, \,\,\mbox{for }\,i= \mbox{1/2 and 3/2}\,.
\end{eqnarray}

These quantities are not sensitive to the branching ratios or total
widths of the resonances. This can be easily seen by inserting Eq.
(A1) into Eq. (A4),

\begin{eqnarray}
\xi^{\ell+}_\frac{1}{2} &=& -C_m \sqrt{2\pi(\ell+1)}\widetilde{A}_{\ell+}\,,\nonumber\\
\xi^{\ell-}_\frac{1}{2} &=& C_m \sqrt{2\pi(\ell)}\widetilde{A}_{\ell-}\,,\nonumber\\
\xi^{\ell+}_\frac{3}{2} &=& C_m \sqrt{\frac{\pi}{2}\ell(\ell+1)(\ell+2)}
        \widetilde{B}_{\ell+}\nonumber\,,\\
\xi^{\ell-}_\frac{3}{2} &=& -C_m \sqrt{\frac{\pi}{2}(\ell-1)\ell(\ell+1)}
        \widetilde{B}_{\ell-}\,,
\end{eqnarray}
where only the photoproduction amplitudes at resonance enter.

\newpage

\begin{table}[htbp]
    \caption{Photon couplings for proton targets and multipolarities
      for $N^*$ Resonances. The numbers are the averaged values from
      PDG98 \protect\cite{PDG98} and Ref. \protect\cite{Krus97}
      (*).}
  \begin{center}
    \leavevmode
    \begin{tabular}{ccccl}
    $N^*$ Resonance & $\Gamma_R [MeV]$ & helicity & $A_{1/2},
          A_{3/2}  [10^{-3}GeV^{-1/2}]$ & 
          $A_{\ell\pm}, B_{\ell\pm}$ \\
    \hline
    $P_{11}(1440)$ & $350\pm 100$ & 1/2 & $ -65\pm 4$  & $A_{1-}=M_{1-}$ \\
    $D_{13}(1520)$ & $120^{+15}_{-10}$ & 1/2 & $ -24\pm 9$  & $A_{2-}=
          \frac{1}{2}(3M_{2-}-E_{2-})$ \\
                 &  & 3/2 & $+166\pm 5$  & $B_{2-}=E_{2-}+M_{2-}$ \\
    $S_{11}(1535)$ & $150^{+100}_{-50} $ & 1/2 & $ +90\pm 30$  &
          $A_{0+}= E_{0+}$ \\
                 & $212\pm 20^*$ &   & $+120\pm 11^*$ &  \\
    $S_{11}(1650)$ & $150^{+40}_{-5}$ & 1/2 & $ +53\pm 16$  & $A_{0+}=
          E_{0+}$ \\
    $D_{15}(1675)$ & $140^{+30}_{-10}$ & 1/2 & $ +19\pm 8$  & $A_{2+}=
          2E_{2+}+M_{2+}$ \\
                 &  & 3/2 & $ +15\pm 9$  & $B_{2+}=E_{2+}-M_{2+}$ \\
    $F_{15}(1680)$ & $130\pm 10$ & 1/2 & $ -15\pm 6$  &
          $A_{3-}=2M_{3-}-E_{3-}$ \\
                &   & 3/2 & $+133\pm 12$  & $B_{3-}=E_{3-}+M_{3-}$ \\
    \end{tabular}
    \label{tab1}
  \end{center}
\end{table}

\begin{table}
    \caption{Parameterization and fitted values from our energy-dependent analysis
      for the coefficients a, b, c, d, e, f and g. Each of the
      coefficients is parameterized in the form $\sum_i \alpha_i
      (q_\eta^{cm}/m_\eta)^{n_i}$ with up to 3 terms. The coefficients
      a, b, c have the dimension of a cross section [$\mu$b], all
      others are dimensionless.}
  \begin{center}
    \begin{tabular}{ccccccc}
         & $n_1$ & $\alpha_1$ & $n_2$ & $\alpha_2$ & $n_3$ & $\alpha_3$\\
      \hline
        a&0& 4.59 $\pm$ 0.04 &2&-9.33 $\pm$ 0.52 & - & -\\
        b&1&-0.29 $\pm$ 0.07 &-& - &-& -\\
        c&2&-4.82 $\pm$ 0.46 &-& - &-& -\\
        d&1& 0.16 $\pm$ 0.03 &-& - &-& -\\
        e&2& 9.6  $\pm$ 2.3  &3&-38.3 $\pm$ 10.  &4& 36.4 $\pm$ 10.  \\
        f&2& 1.70 $\pm$ 0.10 &4&-1.38 $\pm$ 0.32 &-&-\\
        g&6&-10.4 $\pm$ 2.8  &8& 30.7 $\pm$ 6.4 &-&-\\
    \end{tabular}
  \end{center}
  \label{tab2}
\end{table}

\begin{table}[htbp]
    \caption{Resonance properties determined by our analysis. The
          first line for each resonance is taken from the Particle
      Data Tables 1998 \protect\cite{PDG98}. Our analyses (case 1 and
      2) are based on the average values from PDG for the mass $M_R$,
      full width $\Gamma_R$ and $A_\protect{3/2}$, see
      Tab. \protect\ref{tab2} .}
  \begin{center}
    \leavevmode
    \begin{tabular}{rccccc}
    $N^*$ Resonance & $b_\eta=\Gamma_\eta/\Gamma_R$ & $A_{1/2}$ &
          $A_{3/2}/A_{1/2}$ & 
          $\xi_{3/2}$ & $\xi_{1/2}$\\
    & [\% ] & $ [10^{-3}/\sqrt{GeV}]$ & & $ [10^{-4}/MeV]$ & $
          [10^{-4}/MeV]$\\ 
    \hline
    $D_{13}(1520)$ & - & $ -24\pm 9$  & $-6.9\pm 2.6$ & - & - \\
            case 1 & $0.08\pm 0.01$ & $-79\pm 9$ & $-2.1\pm 0.2$ & $0.185\pm
          0.018$ & $-0.087\pm 0.013$ \\
            case 2 & $0.05\pm 0.02$ & $-79\pm 9$ & $-2.1\pm 0.2$ & $0.134\pm
          0.018$ & $-0.087\pm 0.013$ \\
   Ref.\,\cite{mathur98} & - & $-79\pm 9$ & $-2.5\pm 0.2\pm 0.4$ & $0.167\pm
          0.017$ & $-0.066\pm 0.008$ \\
   Ref.\,\cite{Kon80} & 0.09 & -23 & -5.56 & - & -  \\
    \hline
    $F_{15}(1680)$ & - & $ -15\pm 6$  & $-8.9\pm 3.6$ & - & - \\
            case 2 & $0.15^{+0.35}_{-0.10} $ & - & - & $0.16\pm 0.04$
          & - \\
   Ref.\,\cite{Kon80} & 0.8 & 0.0 & - & - & -  \\
    \end{tabular}
    \label{tab3}
  \end{center}
\end{table}

\begin{figure}[htbp]
\centerline{\psfig{file=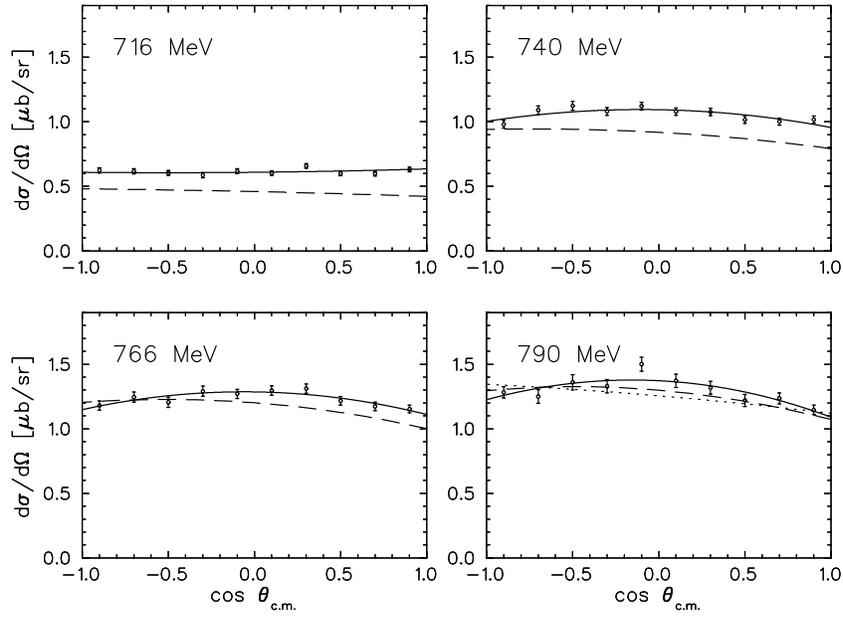,width=11cm,angle=90,silent=}}
\vspace{0.5cm}
\caption[thanks]{\label{fig1} Differential cross sections for 
  $p(\gamma,\eta)p$ at various photon $lab$ energies $E_\gamma^{lab}$.
  The solid lines show our fit to the experimental data of Krusche et
  al. \cite{Krus95}. The dashed lines are the values of the isobar
  model of Kn\"ochlein et al. \cite{Knoe95} and the dotted line at
  $E_\gamma^{lab}=790 MeV$ is obtained from this model if the $D_{13}$
  resonance is turned off. }
\end{figure}
\begin{figure}[htbp]
\centerline{\psfig{file=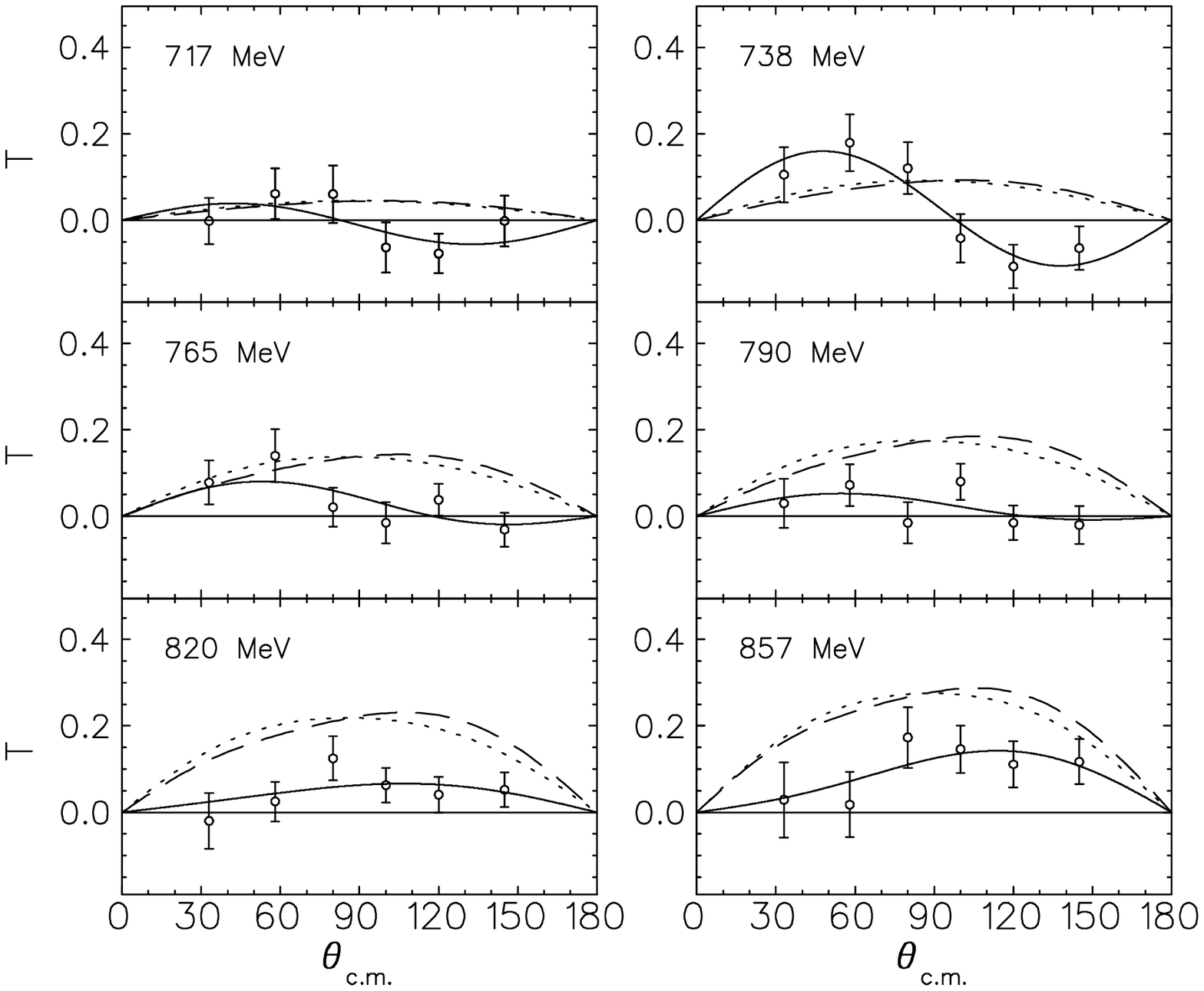,width=12cm,silent=}}
\vspace{0.5cm}
\caption[thanks]{\label{fig2} Target asymmetries for
  $p(\gamma,\eta)p$ at various photon $lab$ energies $E_\gamma^{lab}$.
  The dashed and dotted lines show calculations in the isobar model of
  Kn\"ochlein et al. \protect\cite{Knoe95} with and without the
  $D_{13}(1520)$ resonance. The solid line is the result of our fit to
  the experimental data of Bock et al. \protect\cite{Bock98}. }
\end{figure}
\begin{figure}[htbp]
\centerline{\psfig{file=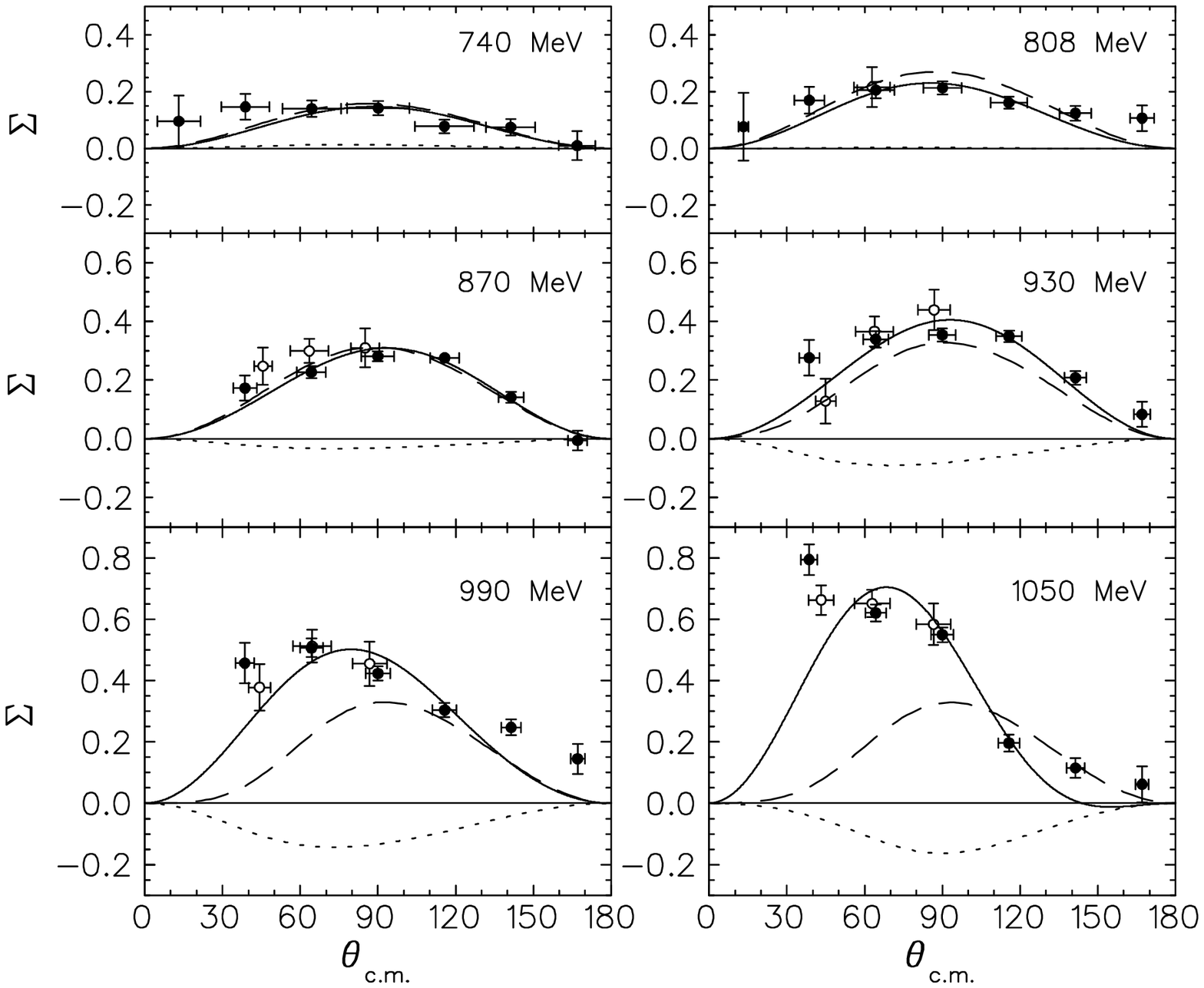,width=12cm,silent=}}
\vspace{0.5cm}
\caption[thanks]{\label{fig3} Photon
  asymmetries for $p(\gamma,\eta)p$ at various photon $lab$ energies
  $E_\gamma^{lab}$. The dashed and dotted lines show calculations in
  the isobar model Kn\"ochlein et al. \protect\cite{Knoe95} with and
  without the $D_{13}(1520)$ resonance. The solid line is the result
  of our fit to the experimental data of Ajaka et al.
  \protect\cite{Hour98}. }
\end{figure}

\begin{figure}[tbp]
\centerline{\psfig{file=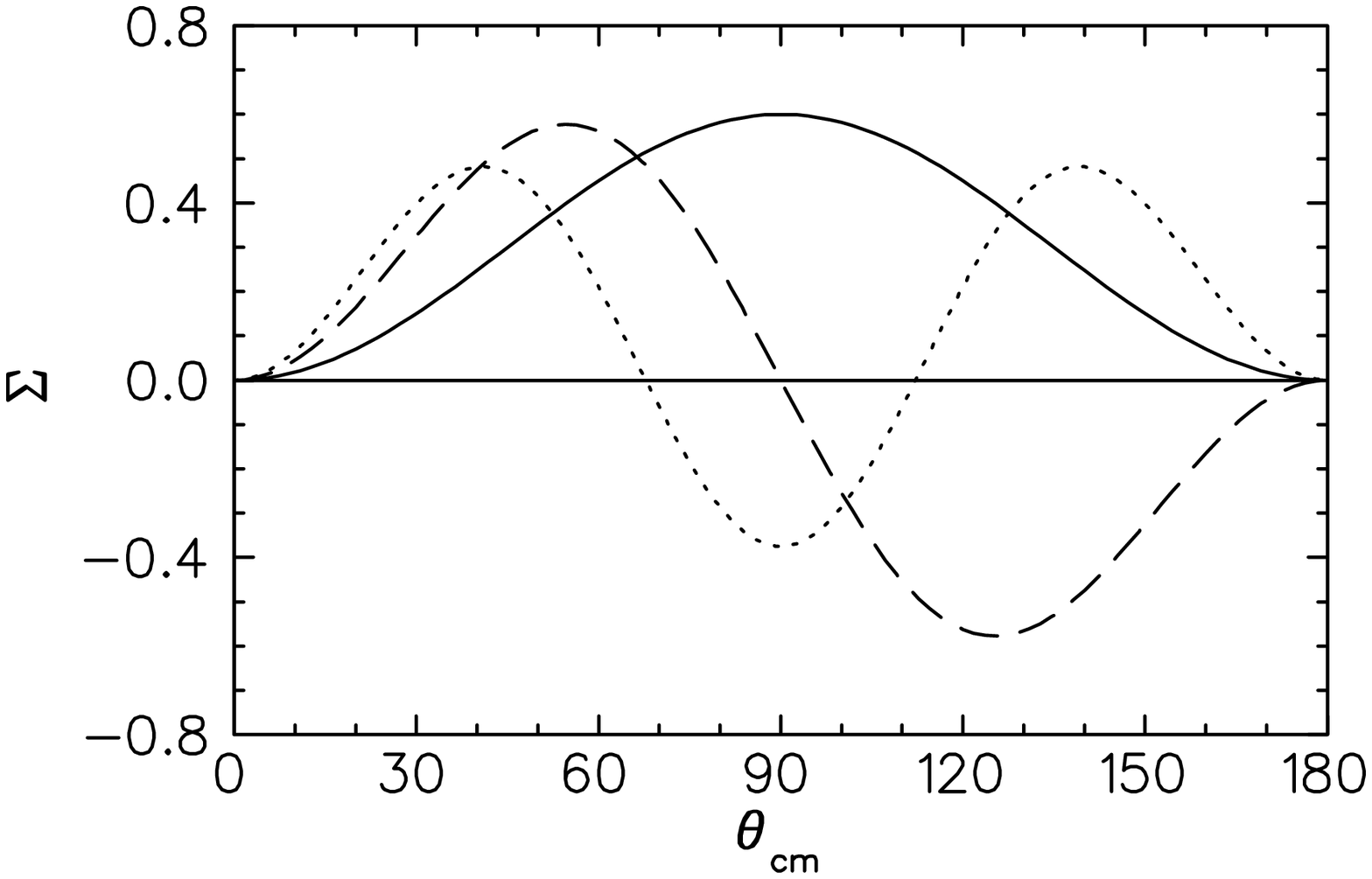,width=7.5cm,silent=}
\psfig{file=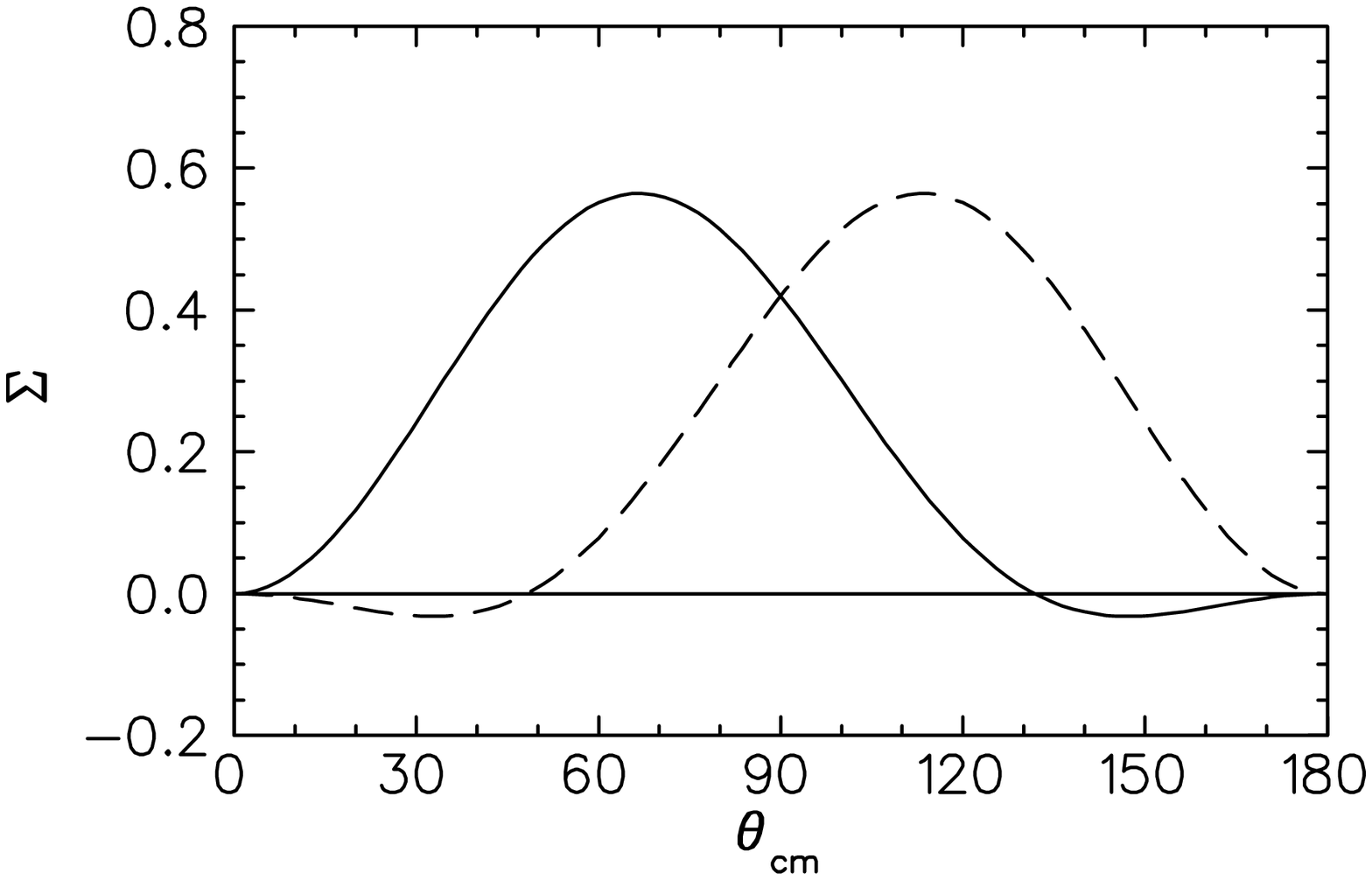,width=7.5cm,silent=}}
\vspace{0.5cm}
\caption[thanks]{\label{fig4} Possible signatures of $N^*$ resonances
  in the photon asymmetry of eta photoproduction. The solid, dashed
  and dotted lines in the left figure show the angular distributions
  for the interference of the dominant $S_{11}$ channel with an
  isolated $d$, $f$, or $g$ wave, respectively. Opposite signs are
  obtained if the photon or eta couplings of the higher resonances
  carry an opposite sign, see Table I. On the right, the situation of
  two overlapping resonances is demonstrated for a ($D_{13}$,
  $F_{15}$) pair (solid curve) and a ($D_{13}$, $F_{17}$) pair (dashed
  curve). }
\end{figure}

\begin{figure}[htbp]
\centerline{\psfig{file=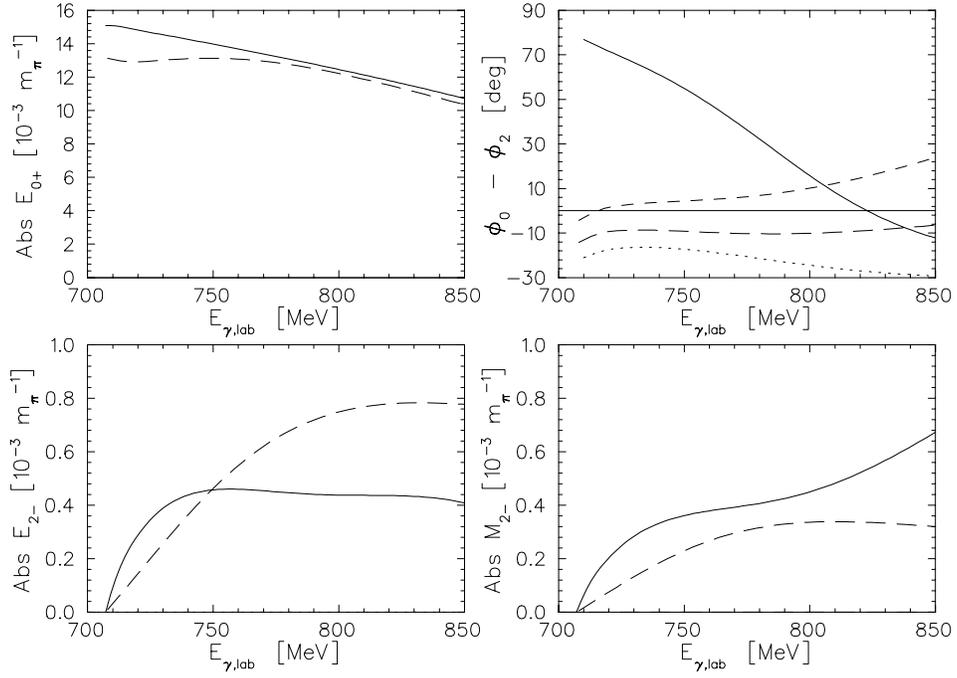,width=12.5cm,angle=90,silent=}}
\vspace{0.5cm}
\caption[thanks]{\label{fig5} Result of the multipole analysis for
  $s$ and $d$ waves. The solid lines show the result of our fit.  The
  short and long dashed lines are obtained using the isobar model of
  Ref. \cite{Knoe95}. In the upper right figure we compare the phase
  difference of our fit with the isobar model. The short and long
  dashed curves show $\phi_{E_{0+}}-\phi_{E_{2-}}$ and
  $\phi_{E_{0+}}-\phi_{M_{2-}}$, respectively, and the dotted line is
  the phase difference for two Breit-Wigner forms. }
\end{figure}

\newpage
\begin{figure}[tbp]
\centerline{\psfig{file=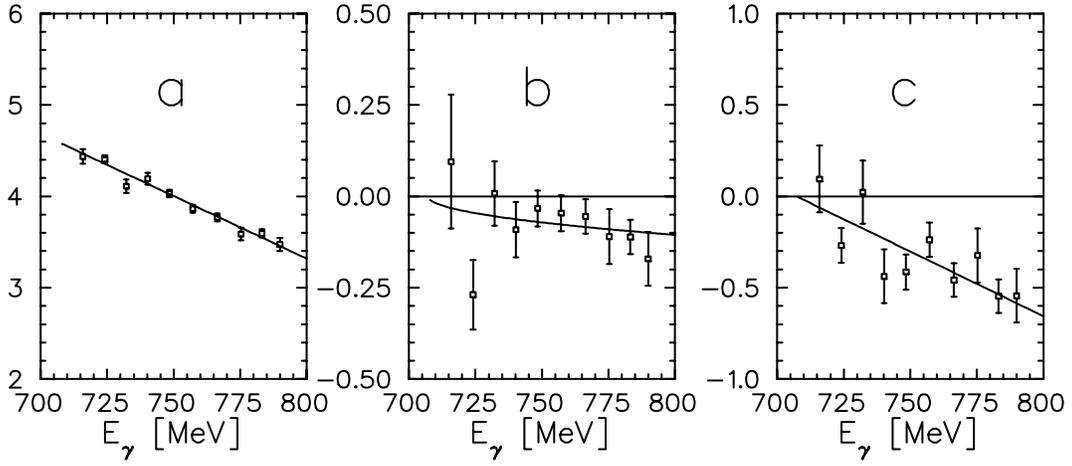,width=14cm,angle=90,silent=}}
\vspace{0.5cm}
\caption[thanks]{\label{fig6} Fit coefficients $a$, $b$ and $c$
  for the differential cross section measured at Mainz \cite{Krus95}. The
  solid lines show polynomial fits allowing for an energy dependent
  solution. The coefficients are given in units of $\mu b/sr$.}
\end{figure}
\begin{figure}[tbp]
\centerline{\psfig{file=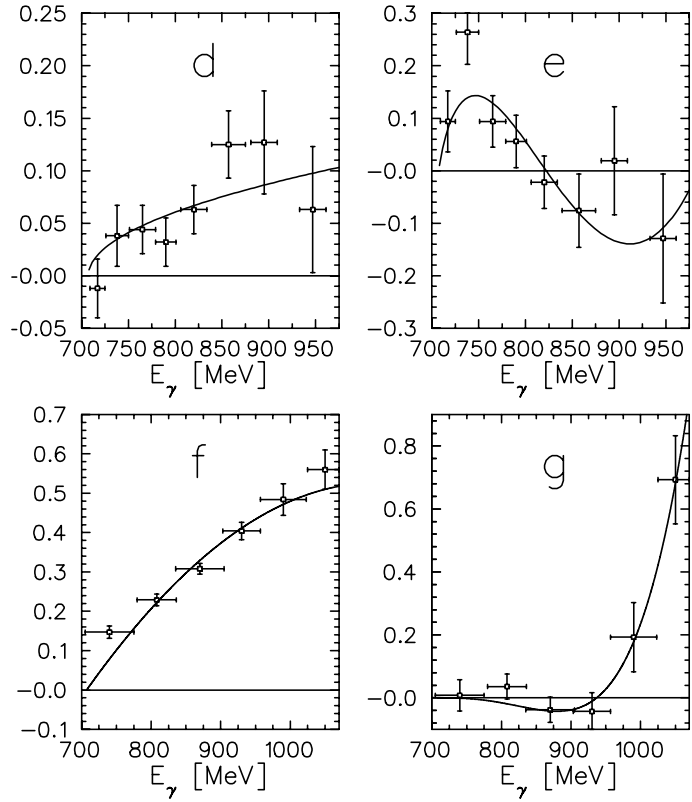,width=9cm,silent=}}
\vspace{0.5cm}
\caption[thanks]{\label{fig7} Fit coefficients $d$, $e$, $f$ and $g$
  of the target and beam asymmetry measured at Bonn
  \protect\cite{Bock98} and GRAAL \protect\cite{Hour98}. The solid
  lines show polynomial fits allowing for an energy dependent
  solution. The coefficients are dimensionless.}
\end{figure}

\end{document}